# Molecular Precursors-Induced Surface Reconstruction at Graphene/Pt(111) Interfaces

Qian Wang, Rui Pang, and Xingqiang Shi*

*Department of Physics, South University of Science and Technology of China, Shenzhen 518055, China*

*E-mail: shixq@sustc.edu.cn

Inspired by experimental observations of Pt(111) surfaces reconstruction at the Pt/graphene (Gr) interfaces with ordered vacancy networks in the outermost Pt layer, the mechanism of the surface reconstruction is investigated by van-der-Waals-corrected density functional theory in combination with particle-swarm optimization algorithm and *ab initio* atomistic thermodynamics. Our global structural search finds a more stable reconstructed (Rec) structure than that was reported before. With correction for vacancy formation energy, we demonstrate that the experimental observed surface reconstruction occurred at the earlier stages of graphene formation: 1) reconstruction occurred when $C_{60}$ adsorption (before decomposition to form graphene) for $C_{60}$ as a molecular precursor, or 2) reconstruction occurred when there were (partial) hydrogens retain in the adsorbed carbon structures for $C_2H_4$ and $C_{60}H_{30}$ as precursors. The reason can be attributed to that the energy gain, from the strengthened Pt-C bonding for C of $C_{60}$ or for C with partial H, compensates the energy cost of formation surface vacancies and makes the reconstruction feasible, especially at elevated temperatures. In the Rec structure, two Pt-C covalent bonds are formed per unit cell, which have a great impact on the adsorbed Gr electronic structures.

# I. INTRODUCTION

The investigations of hybrid organic-metal interfaces are experiencing an explosive growth,[1-4] especially for epitaxial growth of graphene (Gr) layers on metal surfaces. Extraordinary properties such as ambipolar electric field effect, quantization of conductivity, and ultra-high electron mobility of graphene are demonstrated,[5-7] which can be attributed to the unique electronic structure of graphene and its interaction with metal substrates. One of the practical needs is preparing graphene layers on metal surfaces with different thicknesses.[8-9] Although there are many applications for these systems, mechanisms of graphene-metal bonding at their interfaces in the atomic detail are far from complete understanding. Different types of interactions coexist at graphene-metal interfaces, which include chemical bonding, van der Waals (vdW) interaction, Pauli repulsion,[10-13] and the combination of them. For metal substrates whose lattice constants match with that of Gr, strong covalent interactions exist between Gr and metals, and the strong interaction usually destroy the Dirac cone of Gr.[14-15] In the cases of Gr physically adsorbed on or ionically bonded with metal surfaces, the Dirac cone is preserved in the band structure and the energy position of the Dirac cone relative to Fermi energy can be tuned by charge transfer.[16-18]

Surface reconstruction at the Gr/metal interfaces makes it a challenge to understand the Gr-metal contact mechanism.[19-21] Three questions remain to be answered: how is every atom arranged at the interface, can the reconstructed structure be energetically favored, and whether the metal surface reconstruction is induced by Gr adsorption or by its molecular precursors? Fullerene $C_{60}$, a molecular precursor for epitaxial growth of Gr layers, is a typical organic molecule that can reconstruct metal surfaces.[22-23] Although the interactions between graphene-metal surfaces are less strong than that between $C_{60}$-metal, epitaxial growth of graphene induced Pt(111) and Cu(100) surfaces reconstruction were observed.[19-20, 24] However, the studies of the mechanisms of surface reconstruction at Gr/metal interfaces, the global minimum atomic structures at the reconstructed interfaces, and effects of reconstruction on graphene electronic properties are lacking.

In the current work, the mechanism of epitaxial growth of graphene induced Pt(111) surface reconstruction was probed. With the help of particle-swarm optimization algorithm, we found a reconstructed adsorption structure more stable than that proposed before.[19] Then we analyzed the mechanism of reconstruction through energetic considerations, structural analysis, and thermodynamics calculations. Finally, the effects of reconstruction on the electronic structure of the adsorbed graphene were presented.

## II. THEORETICAL METHODS

The Pt(111)-($\sqrt{3} \times \sqrt{3}$)R30°/Gr-(2x2) commensurate structures with reconstruction were searched by using particle swarm optimization methodology as implemented in the CALYPSO code.[25-26] CALYPSO has made successful predictions for surface reconstructions.[27-28] The automatic surface-structure-searching method employed structures swarm intelligence.[28] The fixed bulk region contained three layers to reserve the bulk nature of the substrate; one layer of the outermost Pt atoms and eight carbon atoms were subjected to structure swarm evolution. Once the initial structures were generated, their geometries were optimized by DFT and their total energies were obtained. The structures then evolved towards lower-energy structures, locally and globally, through self- and swarm-structure learning.[28] In this way, global minimum was achieved finally.

DFT calculations with the van der Waals density functional (vdW-DF)[29-30] were employed in the structural optimization and electronic structure calculations. The Vienna ab initio simulation package (VASP)[31-32] was used with Projector Augmented Wave (PAW) potentials.[33] We considered the influence of vdW interactions by using the optB86b-vdW-DF,[34] which offered a good performance when compared to experiments for both chemisorption and physisorption.[11] A kinetic energy cutoff of 500 eV was used. The surface Brillouin zone was sampled with Monkhorst-Pack $k$-meshes of 24 × 24 for the Gr-(2x2)/Pt(111)-($\sqrt{3}\times\sqrt{3}$)R30° surface cell.[35-36] The surface slab contained seven Pt layers.

We adopted *ab initio* atomistic thermodynamics[37-38] to compare the relative stability between adsorption structures with and without surface reconstruction. The adsorption Gibbs free energies of graphene adsorbed on Unrec (abbreviation of unreconstructed) and Rec (reconstructed) Pt(111) surfaces, $G_{\text{Gr/Pt(Unrec)}}^{\text{ads}}$ and $G_{\text{Gr/Pt(Rec)}}^{\text{ads}}$ are given as the following, respectively:

$$G_{\text{Gr/Pt(Unrec)}}^{\text{ads}}(T,p,N_{\text{Pt}},N_{\text{Gr}}) = G_{\text{Gr/Pt(Unrec)}}(T,p) - G_{\text{Pt(Unrec)}}(T,p) - G_{\text{Gr}}(T,p), \quad (1)$$

$$G_{\text{Gr/Pt(Rec)}}^{\text{ads}}(T,p,N'_{\text{Pt}},N_{\text{Gr}}) = G_{\text{Gr/Pt(Rec)}}(T,p) - G_{\text{Pt(Rec)}}(T,p) - G_{\text{Gr}}(T,p). \quad (2)$$

Here $G_{\text{Gr/Pt(Unrec)}}$, $G_{\text{Pt(Unrec)}}$, and $G_{\text{Gr}}$ are the adsorption Gibbs free energies of Unrec Gr/Pt system, pure Unrec Pt surface, and free standing graphene, respectively. Similarly definitions are used for the Rec case. The difference of adsorption Gibbs free energies between Rec and Unrec structures, $\Delta G^{ads}$ is calculated by the subtraction of these above two free energies:

$$\Delta G^{ads}(T,p) = G_{Gr/Pt(Rec)}(T,p) - G_{Pt(Rec)}(T,p) - [G_{Gr/Pt(Unrec)}(T,p) - G_{Pt(Unrec)}(T,p)]. \quad (3)$$

The vacancy formation Gibbs free energy due to the missing of surface Pt atoms, $\Delta G^{vac}$ is given as:

$$\Delta G^{vac}(T,p) = G_{Pt(Rec)}(T,p) + \mu_{Pt}(T,p_{Pt}) - G_{Pt(Unrec)}(T,p). \quad (4)$$

The chemical potential of the missing Pt atom $\mu_{Pt}$ is given by:

$$\mu_{Pt}(T,p_{Pt}) = E_{Pt}^{bulk} + \tilde{\mu}_{Pt}(T,p^0) + k_B T \ln\left(\frac{p_{Pt}}{p^0}\right), \quad (5)$$

where $E_{Pt}^{bulk}$ is the total energy of a Pt atom in bulk phase, and for this we use our DFT result; and $\tilde{\mu}_{Pt}(T,p^0)$ is the standard chemical potential including all entropy contributions. For simplicity, the $\tilde{\mu}_{Pt}(T,p^0)$ values from the JANAF thermodynamic tables are used.[39-40]

By adding $\Delta G^{ads}$ and $\Delta G^{vac}$, the variation of the free energy between Rec and Unrec can be obtained:

$$\Delta G = \Delta G^{ads}(T,p) + \Delta G^{vac}. \quad (6)$$

The Gibbs free energy can be described as

$$G(T,p) = E^{tot} + F^{vib} - TS^{conf} + pV. \quad (7)$$

The total energy $E^{tot}$ plays the major role in Gibbs free energy and can be achieved from DFT calculations, $F^{vib}$ accounts for the vibrational contribution, $S^{conf}$ is the configurational entropy. The differential Gibbs free energy in eq. (3) can be replaced by DFT calculated total energies, which is discussed in Refs..[39, 41]

### III. RESULTS AND DISCUSSIONS

#### A. Structural search

Since the adsorption energy increase of the Rec structure as proposed in Ref.[19] relative to the Unrec one cannot offset the Pt(111) surface vacancy formation energy of 1.18eV, CALYPSO method was used to find the most stable reconstructed structure. The unreconstructed structure was also searched by CALYPSO. The most stable unreconstructed (Unrec) and reconstructed (Rec1) structures are shown in Figure 1(a) and (b). Figure 1(c) shows the second most stable reconstructed structure (Rec2) which is same to the model proposed in Ref..[19] For both Rec1 and Rec2 structures, one third of surface Pt atoms are missing. These two structures are similar to each other: from Rec2, the graphene layer in Rec1 is just moved along the surface [110] direction by a C-C bond length (compare Figure 1b and c). The adsorption energies and the nearest distances between carbon and platinum are shown in Table 1. The nearest distance between carbon and platinum in the Unrec structure is about 3.3 Å, which is in the range of a typical physisorption; while Rec1 and Rec2 are chemical adsorption of graphene on Pt(111) surface with the Pt-C distances of about 2.3 Å. Within a ($\sqrt{3} \times \sqrt{3}$) R30° cell two bonds are formed in Rec1, and one bond is formed in Rec2. Rec1 has the largest adsorption energy. With respect to Unrec structure, the adsorption energies of two reconstructed structures have increased: the increment of Rec1 structure is 0.20eV, while the increment of Rec2 structure is 0.02eV from vdW-DF results. This can attribute to the chemical bonds between carbon atoms and platinum atoms. More chemical bonds are formed in Rec1, and hence Rec1 is more stable than Rec2.

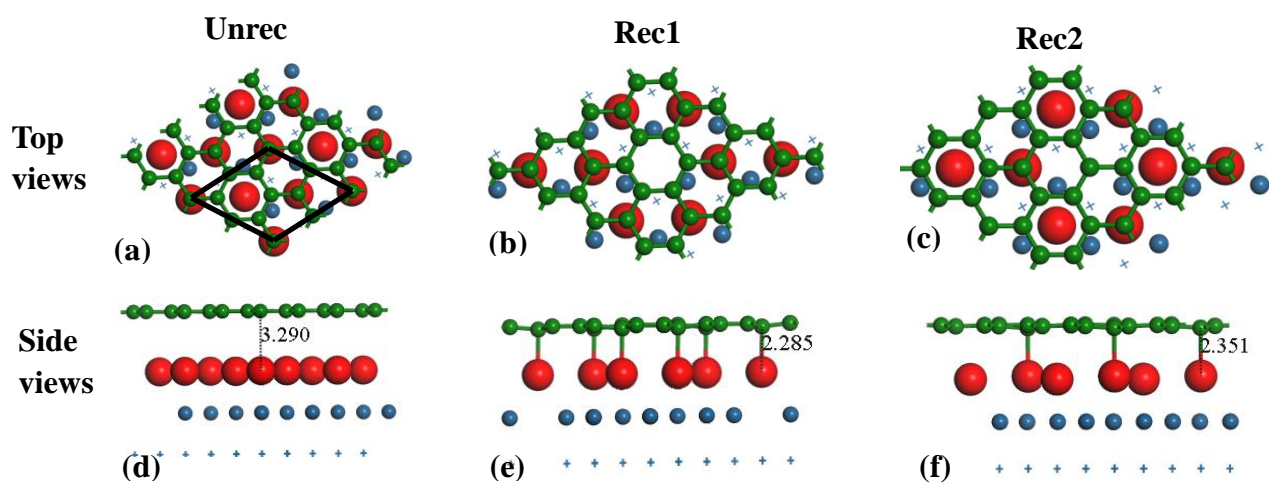

Figure 1. (Color online) Graphene on Pt(111): top and side views of the most stable unreconstructed structure(Unrec), most stable reconstructed structure (Rec1), and second most stable reconstructed structure (Rec2); Rec2 is the model proposed in Ref..[19] Green balls represent carbon atoms; red balls, blue balls, and blue crosses represent the first, second, and third layers of Pt atoms, respectively. Only the top three Pt layers are shown. The surface unit cell is shown with a rhombus in panel (a).

Table 1. Adsorption energies per unit cell and the nearest C−Pt distances $d$ from vdW-DF calculations. The vdW-DF energy is decomposed into $E_c^{nl}$ and the remainder ($E^r$ for short), see text for details. Negative (positive) energies mean binding (unbinding). For comparison to Ref.,[19] their local density approximation (LDA) results are also presented.

|  | $E_{ads}$(vdW-DF) (eV) | $E^r$ (eV) | $E_c^{nl}$ (eV) | $d$(vdW-DF) (Å) | $E_{ads}$(LDA) (eV) | $d$(LDA) (Å) |
|---|---|---|---|---|---|---|
| Unrec | -0.66 | 0.43 | -1.09 | 3.29 | -0.33(-0.3[a]) | 3.47(3.30[a]) |
| Rec1 | -0.86 | 0.80 | -1.66 | 2.29 | -0.97 | 2.22 |
| Rec2 | -0.68 | 0.90 | -1.58 | 2.35 | -0.74(-0.6[b]) | 2.23(2.28[b]) |

[a]Reference;[36]  [b]Reference.[19]

Our calculated results with local density approximation (LDA) for adsorption energies of Unrec and Rec2 are -0.33 and -0.74eV, resp., which are consistent with those in Refs.[19, 36] (See Table 1). The results from vdW-DF are quite different from that calculated by LDA -- the difference in adsorption energies between Rec2 and Unrec is much smaller in the vdW-DF results. This indicates that LDA is not adequate to describe the relative stability between physisorption (Unrec) and chemisorption (Rec2). To probe why vdW-DF and LDA results are different, we decompose the vdW-DF energy in to two parts: the *nonlocal* electron correction energy $E_c^{nl}$ as given in Dion *et al*[29] and the remainder $E^r$, namely $E_{xc}^{vdW-DF} = E_x^{GGA} + E_c^{LDA} + E_c^{nl} = E^r + E_c^{nl}$.[34] Here $E_x^{GGA}$ is the exchange energy with a generalized-gradient approximation (GGA) functional, $E_c^{LDA}$ is the LDA correlation energy. The nonlocal term $E_c^{nl}$ gives a similar energy difference to that given by LDA between Unrec and Rec2 structures (Table 1), while the remainder $E^r$ gives unbound results and gives an inversed energy difference to that given by LDA: this is the reason why vdw-DF gives smaller energy difference than that given in LDA.

The vacancy formation energy of missing one Pt atom per unit cell is 1.18 eV from vdW-DF, while the adsorption energy increase is only 0.20eV from Unrec to Rec1 structure (Table 1, vdW-DF results). Obviously, the energy gain in adsorption energy cannot compensate the vacancy formation energy. That is, reconstruction is difficult to occur through graphene adsorption. According to the fact that forming more Pt-C chemical bonds enhances the adsorption energy, and considering the chemical properties of molecular precursors for the growth of graphene, namely the molecular precursors ethylene and $C_{60}$ binds stronger with Pt(111) than graphene does. We wondered if the reconstructed structures might be induced by the molecular precursors in the earlier stages of graphene formation. The $C_{60}$ precursor binds stronger with Pt(111) than Gr does (will be discussions later), and it is well accepted that $C_{60}$ can reconstruct Pt(111) and other close-packed metal surfaces.[22-23, 42-44] However, for $C_{60}$ reconstructed Pt(111), at most one-twelfth (~8%) surface Pt atoms are missing,[42] which ratio agrees well to that observed in the scanning tunneling microscope (STM) images after Gr formation[19] -- the Pt(111)- ($\sqrt{3} \times \sqrt{3}$) reconstructed areas are covered up to 20% area of the sample [or $(1/3) \times 20\% \cong 7\%$ of the sample] and this area has never been found isolated from the unreconstructed parts.[19] This means that the ~8% vacancies induced by $C_{60}$ adsorption gathered together to form $\sqrt{3}\times\sqrt{3}$ areas after Gr formation, presumably due to only in this way Gr can form chemical bonds with Pt(111) in the $\sqrt{3}\times\sqrt{3}$ areas.

In experiment, thermal decomposition of organic molecules were used for epitaxial growth of graphene;[19] and three molecular precursors were used, including ethylene, planar $C_{60}H_{30}$,[45] and spherical $C_{60}$. During evaporating, the substrate temperature was chosen as the minimum temperature for molecular decomposition (or low Gr formation temperatures), and long evaporating time up to one hour was employed. Under these experimental conditions, the molecular precursors can reconstruct the Pt(111) surface with platinum atoms dug out: experimental evidences have been demonstrated that Pt atoms are removed from the surface and diffuse towards the step edges, leaving a vacancy behind upon annealing the molecular precursors.[19, 21] Furthermore, during growth, a strong graphene−metal interaction was indicated.[9] It can be inferred that reconstruction could be occurred in the early stage of graphene formation.

Nonplanar bended graphene, such as Maze-like[24] and dome-shaped[46] graphene, can form more and stronger chemical bonds with metal surfaces; but graphene itself cannot bend. Considering the hydrogen-containing molecular precursors, we wondered if partial hydrogen retain in the carbon structures could bend the planar graphene and enhance its bonding strength to Pt, which is confirmed in Ref..[47] In view of the above types of precursors, the adsorption of hydrogenated graphene (H-Gr), ethylene, and benzene (a simplification of planar $C_{60}H_{30}$) were considered. In the following, we use H-Gr adsorption as an example to clarify our point that the carbon structures with (partial) hydrogens make Pt surface reconstruction easier than that in pure Gr adsorption (see Figure 2 and Table 2). Ethylene and benzene adsorptions show the same trend (Table 2 and Figure 3).

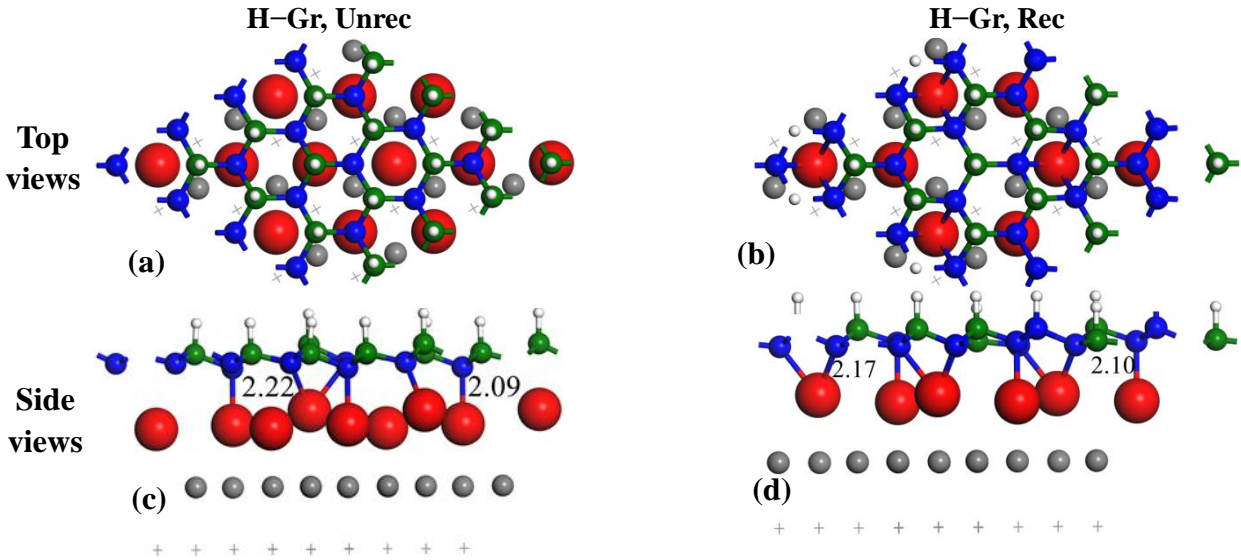

Figure 2. (Color online) H−Gr on Pt(111): top and side views of unreconstructed and reconstructed adsorption structures of H−Gr/Pt. Red balls, gray balls, and gray crosses represent the first, second, and third layer Pt atoms, respectively; and white balls represent hydrogen atoms. Carbon atoms in H-Gr are divided into two layers, wherein green balls represent the upper carbon atoms and blue balls represent the lower ones. Only the top three Pt layers are shown.

Table 2. Adsorption energies per cell (or per molecule) and the nearest C−Pt distances from vdW-DF calculations for the unreconstructed and reconstructed H-Gr/Pt(111), $C_2H_4$/Pt(111), and benzene/Pt(111).

|  | H-Gr | | $C_2H_4$ | | Benzene | |
| --- | --- | --- | --- | --- | --- | --- |
|  | Unrec | Rec | Unrec | Rec | Unrec | Rec |
| $E_{ads}$(eV) | -6.03 | -7.12 | -0.80 | -1.50 | -2.29 | -3.31 |
| $d$(Å) | 2.09/2.22 | 2.10/2.17 | 2.17 | 2.13 | 2.16/2.20 | 2.14 |

In the adsorption structures of Gr/Pt(111), the nearest Pt-C distance changes from 3.3 Å in physisorbed Unrec structure to 2.3 Å in chemisorbed Rec structure. The nearest Pt-C distance is smaller in H-Gr/Pt(111) than that in Gr/Pt(111). Since in H-Gr/Pt Unrec and Rec structures, hydrogen promoted the corrugation of graphene. This corrugation changes the carbon bonds from inplane $sp^2$ to out-of-plane $sp^3$-like and strengthens the C-Pt chemical bonds, which increases the adsorption energy and facilitates the occurrence of platinum surface reconstruction. The Unrec H-Gr/Pt(111) structure has four Pt-C bonds per (√3×√3)R30° cell, one with a bond length of 2.09 Å and three with 2.22 Å; while the Rec H-Gr/Pt(111) structure has four Pt-C bonds per (√3×√3)R30° cell, one with 2.10 Å and three with 2.17 Å. Stronger chemical bonding are formed in general between H-Gr and Pt surface in the Rec structure. The adsorption energies of Unrec and Rec H-Gr/Pt(111) are shown in Table 2. The difference in adsorption energy between Unrec and Rec H-Gr/Pt(111) is 1.09eV; compared to the 0.2eV difference for that in Gr/Pt(111), the difference in adsorption energy of 1.09eV brought by H-Gr is much larger and closer to the vacancy formation energy of 1.18eV. Hence reconstruction is more likely to occur in H-Gr/Pt(111) system than in Gr/Pt(111), or more general, reconstruction is more likely to occur for carbon structures with H.

Similarly, we considered the effects of ethylene and benzene adsorption on Pt(111) surface reconstruction. In all cases, partial $sp^3$ hybridization also appeared for the adsorbed molecules with H. In Unrec and Rec structures of $C_2H_4$/Pt(111), the Pt-C bond lengths are 2.17Å and 2.13 Å, respectively (Table 2). The difference of adsorption energy corresponding to $C_2H_4$/Pt(111) is 0.69eV. In benzene/Pt(111) Unrec structure, two types of bonds are formed with four bonds length of 2.21Å and two 2.16 Å, while in Rec structure the bond length is 2.14 Å between benzene six carbons and the six Pt surface atoms surrounding the vacancy. The difference of adsorption energy for benzene/Pt(111) is 1.02eV. So, both $C_2H_4$/Pt(111) and benzene/Pt(111) have the same trend in adsorption energy increase as that in H-Gr/Pt(111), and all the three cases for carbon structures with H make the Pt surface reconstruction easier than that in Gr/Pt(111).

For $C_{60}$ molecules as precursors, the surface reconstruction can occur when $C_{60}$ adsorption (namely before decomposition of $C_{60}$). Compared to Pt-graphene interaction, the interaction of Pt-$C_{60}$ is stronger: the adsorption height and Pt-$C_{60}$ bond length are 1.7 Å and 2.1 Å, respectively;[42] while in Gr/Pt(111) the height and bond length are both 2.3 Å. The origin for this difference in bonding strength is due to the partial $sp^3$ hybridization formed between $C_{60}$ and Pt surface. The similar partial $sp^3$-like bonding exists in the interaction between Pt surface and H-graphene, ethylene, or benzene. In all cases for molecular precursors of $C_{60}$ or of carbon structures with H, the enhanced Pt-C chemical bonds effectively promote the occurrence of Pt surface reconstruction.

### B. Atomistic thermodynamics

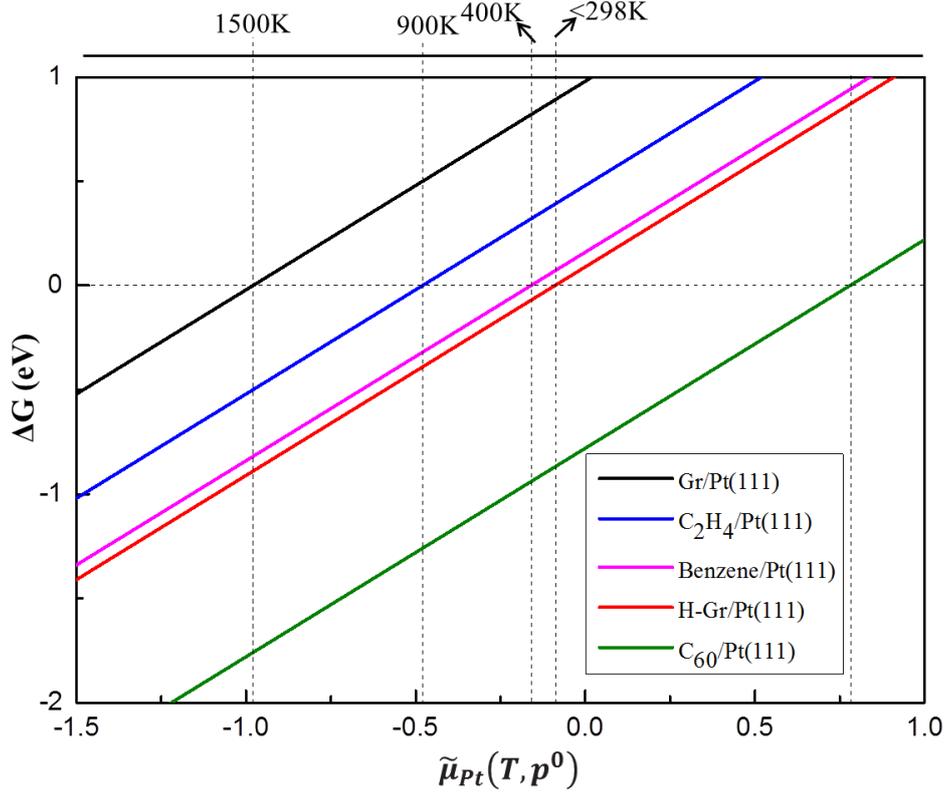

Figure 3. Phase diagram of $\Delta G$ vs $\tilde{\mu}_{Pt}(T,p^0)$. $\Delta G$ is the variation of the Gibbs free energy between Rec and Unrec, and $\tilde{\mu}_{Pt}(T,p^0)$ is the Pt chemical potential as a function of temperature. Black, blue, pink red, and green lines represent $\Delta G$ of Gr/Pt(111), $C_2H_4$/Pt(111), benzene/Pt(111), H-Gr/Pt(111), and $C_{60}$/Pt(111), respectively. The energy of bulk Pt is set as zero of Pt chemical potential.

The formulas for calculating $\Delta G$ are detailed in the Method Section. Reconstruction can be lower in free energy when $\Delta G < 0$. Except for $C_{60}$, for all systems considered in Figure 3, $\Delta G$ are all larger than zero for the Pt chemical potential at zero, namely for the DFT total energy of bulk Pt as a reference. However, H-Gr, ethylene, and benzene adsorption make it easier to reconstruct the Pt(111) surface relative to that of Gr adsorption. And $C_{60}$ adsorption makes the Pt surface reconstruction most easily.

In H-Gr/Pt model, the $\Delta G^{vac}$ and $\Delta G^{ads}$ are 1.18 eV and 1.09 eV respectively, the 0.09 eV energy difference can be overcame at below room temperature (Figure 3). We labeled the temperature at $\Delta G = 0$ for each model system in Figure 3. The temperature of reconstruction in Gr/Pt(111), $C_2H_4$/Pt(111), benzene/Pt(111), and H-Gr/Pt(111) systems is about 1500, 900, 400, and less than 298 K, respectively.

The experimental annealing temperature for the (√3×√3)R30° reconstruction is 900–1000 K, and slow evaporation(up to 1 hour) in ultrahigh vacuum conditions. In such an annealing temperature and long annealing time, Pt surface reconstruction would induce by ethylene $C_2H_4$, $C_{60}H_{30}$ adsorption (or possibly other carbon structures with partial hydrogen like $C_2H_2$ and H-Gr). The reconstruction would take place when $C_{60}$ adsorb with lower annealing temperatures. In summary, the reconstructed ($\sqrt{3}\times\sqrt{3}$)R30°-Pt(111)/2×2-Gr structure can attribute to the adsorption induced by carbon structure with hydrogen, or by fullerenes like $C_{60}$. By adsorbing graphene through molecular precursors of hydrogen-containing carbon structures or $C_{60}$, reconstruction of Pt(111) surface can occur. At the same time, reconstruction produced a strong bonding between graphene and platinum, which changed the electronic structure of graphene.

## C. Electronic structure

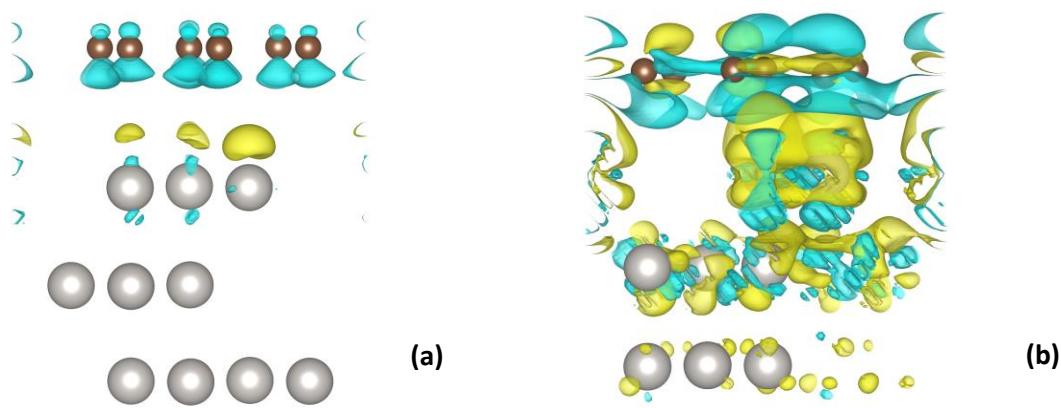

Figure 4. Differential charge density plots of Unrec (a) and Rec1 structures (b). Yellow and blue represent for electron accumulation and depletion, respectively (with isosurface value of 0.001 e/Bohr$^3$).

The differential charge density plots of Unrec and Rec1 structures are shown in Figure 4. According to the differential charge distribution, electrons accumulate at Pt surface and deplete around graphene. In Unrec structure, accumulation and depletion of electrons are less significant, which indicates physical adsorption plays the dominant role. In Rec1 structure, electrons accumulate at the middle between graphene and Pt(111), which indicates the formation of covalent bonds between graphene and Pt surface.

The Bader charge[48] of these two systems were calculated, the results are coincided with the differential charge analysis. Electrons transfer from graphene to Pt in both two systems are small, the amounts of charge transferred in Unrec and Rec1 structures are 0.06 electrons and 0.02, respectively; corresponding a physisorption character and a covalent bonding character, respectively.

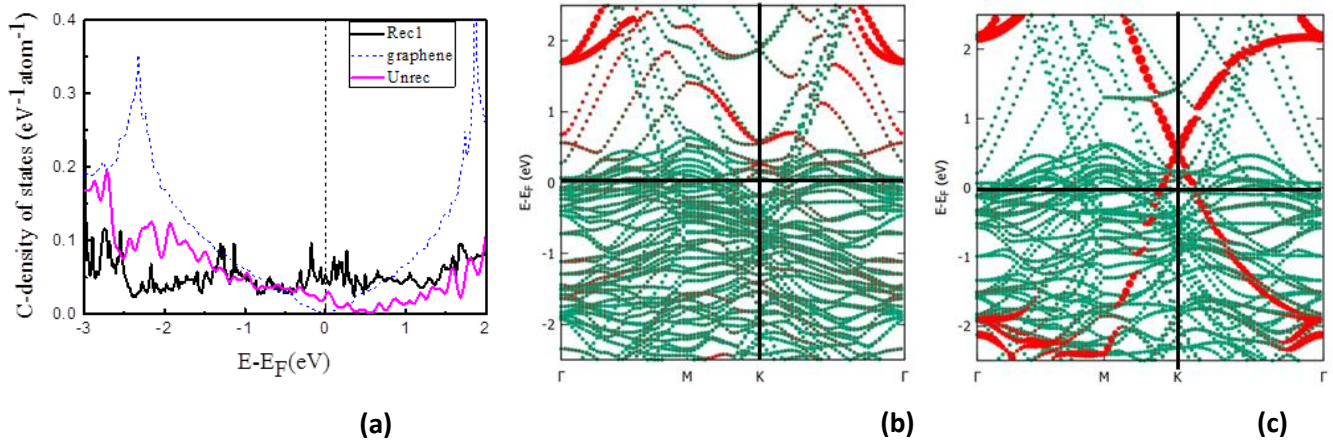

(a) (b) (c)

Figure 5. (a) Partial density states of C in Rec1 of Gr/Pt(111), free standing graphene, and Unrec of Gr/Pt(111). Band structures of Gr/Pt(111) in Rec1 and Unrec structures. The Fermi energy is set to zero. The amount of carbon contribution is indicated by the size of the red dots in the band structure plots.

To better understand the influence of unreconstructed and reconstructed Pt surface on graphene electronic structure, we present the partial density of states (PDOS) plots of C in Rec1 of Gr/Pt(111), free standing graphene, and Unrec of Gr/Pt(111) in Figure 5(a). The vertical dashed line represents the Fermi level. Compare to the free standing graphene, the electronic structure of graphene adsorbed on reconstructed and unreconstructed Pt surface have changed a lot. The electron density of states of adsorbed graphene is broadened. As shown that, the Dirac cone of free standing graphene located at the Fermi level. For graphene on unreconstructed Pt surface, the Dirac cone of graphene remained and located at about 0.5eV above the Fermi level, while the Dirac cone of graphene adsorbed on reconstructed Pt surface are destroyed.

From band structures plots (Figure 5b and c), when graphene is adsorbed on unreconstructed Pt surface, the Dirac cone at *K*-point can still be clearly identified, and the Dirac cone locates at about 0.5e V above Fermi level due to charge transfer, which is consistent with the PDOS and Bader charge results. This illustrates that the interaction between graphene and unreconstructed Pt surface is weak and belongs to physical adsorption. Since the Dirac cone locates at the Fermi level for neutral freestanding graphene, the upwards shifted Dirac cone in Unrec structure should attribute to electron transfer from graphene to Pt surface.[36] For graphene adsorption on the reconstructed Pt surface, dramatic change is occurred on the projected-bands of graphene. The destruction of Dirac cone indicates that there is strong chemical bonding between reconstructed Pt surface and graphene, which agrees with the PDOS and differential charge plots. This character is similar to Gr on Ni(111) and Co(0001) surfaces, whose lattices match the Gr lattice well.[36] Reconstruction makes the Pt-C bonding strong and makes the Pt(111) surface act as the lattice matched metals of Ni(111) and Co(0001).

## IV. CONCLUSIONS

To summarize, firstly, a more stable reconstructed surface structure at the Gr/Pt(111) interface was found in our current work. Based on the considerations of vacancy formation energy, the reconstruction is hard to happen for Gr adsorption. As the solution, we considered $C_{60}$ and hydrogen-containing precursors. For hydrogen-containing carbon structures, we considered $C_2H_4$, benzene, and H-Gr. We showed that $C_2H_4$, benzene, and H-Gr facilitated the reconstruction of Pt(111) surface as well as $C_{60}$, and so reconstruction may happen at the early stages of graphene formation. Further, through atomistic thermodynamics analyses, we described that platinum surface reconstruction could be realized at different annealing temperatures for different molecular precursors. The mechanism of molecular precursor adsorption induced surface reconstruction at the Gr/Pt(111) interface could be applied to epitaxial growth of the graphene on other metal surfaces, e.g. on Cu(100). The electronic structure of graphene changed a lot when adsorbed on the reconstructed Pt(111) surface. Electrons were accumulated in between Pt surface and graphene in the Rec structure due to the formation of covalent bonds between Pt-C. In the Unrec structure The Dirac point of Gr is retained and shifted upwards due to electron transference; while it was destroyed in the Rec structure. Physical and chemical Gr-Pt interactions are in the Unrec and Rec structures, respectively.

## ACKNOWLEDGMENTS

This work is supported by National Natural Science Foundation of China (Grants 11474145 and 11334003). We thank the National Supercomputing Center in Shenzhen for providing computation time.